\documentstyle[aps]{revtex}
\def\be{\begin{equation}}
\def\ee{\end{equation}}
\def\bea{\begin{eqnarray}}
\def\eea{\end{eqnarray}}
\begin{document}  

\title{ A New Doublet-Triplet Splitting Mechanism for Supersymmetric 
SO(10) and Implications for Fermion Masses}

\author{ Z. Chacko and Rabindra N. Mohapatra}
\address{\it Department of Physics, University of Maryland,
College Park, MD 20742, USA}

\maketitle

\begin{abstract}
We present a new mechanism for doublet-triplet splitting 
in supersymmetric SO(10) models using a missing vev pattern which is 
different from the one used in the currently popular Dimopoulos-Wilczek 
method. In our method, the doublets in a {$\bf 16,\bar{16}$} pair are the 
ones split from the rest of the multiplet and are 
then mixed with the doublets from one or two {\bf 10}'s giving rise to the 
doublets {$\bf H_u$} and {$\bf H_d$} of the standard model. This approach
provides a natural way to understand why top quark is so much heavier 
than the bottom quark. It also enables us to generate both 
hierarchical and nonhierarchical pattern for neutrino masses, the latter 
being of interest if neutrino is the hot component of the 
dark matter of the universe. We construct a simple, 
realistic model based on this idea. The model uses only simple 
representations, has no unwanted flat directions and maintains 
coupling constant unification as a prediction.
\end{abstract}

\hskip 4cm UMD-PP-99-031~~~~e-mail address: rmohapat@physics.umd.edu
 
\section{Introduction}\hspace{0.5cm} 
Supersymmetry appears to be the simplest way to provide a 
satisfactory resolution of two of the outstanding puzzles of the standard 
model: (i) the Higgs mass problem and (ii) the origin of electroweak 
symmetry breaking. The minimal supersymmetric extension of the standard 
model (MSSM) which embodies both these features has rightly been the 
focus of intense activity in the past decade. An additional advantage
of the MSSM particle content is that it automatically leads to the 
unification of the gauge couplings near $2\times 10^{16}$ GeV raising the 
hope that the high energy theory of particles and forces may indeed be a 
supersymmetric grand unified theory\cite{1,2} based on some simple group.
One of the currently favoured groups for grand unification is the SO(10)
group which provides a natural way to incorporate the neutrino masses. This
will be the subject of this letter.

A key problem of all SUSY GUTs is how to split the weak MSSM doublets 
from the color triplet fields that accompany them as part of the 
representation of the GUT symmetry. This is essential for constructing 
realistic SUSY GUTs since both coupling constant unification and 
suppression of proton decay require that the MSSM doublets $H_u$ and 
$H_d$ be at the weak scale whereas the triplets which mediate proton 
decay must have GUT scale mass. This is the famous doublet-triplet 
splitting problem (DTS).

The most popular mechanism for implementing the DTS in SO(10) SUSY GUTs
is the missing vev pattern advocated by Dimopoulos and Wilczek(DW)\cite{DW}
and applied to realistic models by Babu and Barr\cite{babu}. Further analysis
and improved applications of this idea have been carried out in 
Ref.\cite{babu1}. The key observation underlying this procedure is 
that the vev
pattern of the {\bf 45}-dim. Higgs multiplet, $A$ of the following form i.e.
$<A>=~i\tau_2\otimes Diag(a,a,a,0,0)$ leads to one pair of massless  
standard model doublets for each of the two {\bf 10}-dim Higgs fields that 
couples to it. After giving mass to two of these four doublets the low 
energy spectrum is the MSSM.

In this paper we present an alternative to the Dimopoulos-Wilczek mechanism
for doublet triplet splitting in SO(10). The idea behind our
method is to use missing vev patterns to split the doublets in a {$\bf 
16,\bar{16}$} pair (denoted by $P$ and $\bar{P}$)  
from the rest of the multiplet. These doublets are 
then mixed with the doublets from a pair of {\bf 10}'s so that the 
{$\bf H_u$} and {$\bf H_d$} of the standard model emerge as linear 
combinations of the {\bf 16}'s and the {\bf 10}'s. This approach 
provides a natural way to understand why the top quark is so much heavier 
than the bottom quark and has all the ingredients needed to generate a 
realistic mass and mixing pattern for the charged fermions. We also point out
how this new doublet-triplet splitting mechanism can be used to generate
nonhierarchical mass pattern for neutrinos using what is called the type 
II seesaw mechanism\cite{goran}. This is to be contrasted with the 
usual seesaw (type I) mechanism\cite{seesaw} which leads to strong hierarchy 
among the neutrino masses. If the neutrinos are to constitute the hot 
dark matter of the universe, then such a nonhierarchical pattern is 
clearly needed. It is our hope that the observations 
made in this paper will open up new ways to build realistic SO(10)
models incorporating desired neutrino mass patterns.

The missing vev patterns required for splitting the doublets in $P$ and 
$\bar P$ from the rest of the multiplets are:

\noindent (i) a {\bf 45} dimensional Higgs field $A$ with a vev pattern
complimentary to the Dimopoulos-Wilczek type ie. $<A>=~i\tau_2\otimes 
Diag (0,0,0,b,b)$ that couples to $P$ and $\bar P$ as $PA\bar{P}$.

\noindent  (ii) another ${\bf 16, \bar{16}}$ pair $C$ and $\bar{C}$ with
vevs along the right handed neutrino direction that couple to $P$ and 
$\bar P$ as $\bar{C}\bar{A}P + C\bar{A}\bar{P}$ where $\bar{A}$ is 
another {\bf 45} which does not have any vacuum expectation value.

In order to understand how this works it is best to decompose the 
representations according to their transformation properties under 
$SU(4)_C \times SU(2)_L \times SU(2)_R$. Then
\begin{eqnarray}
{\bf 45} &=& {\bf (15,1,1) + (6,2,2) + (1,3,1) + (1,1,3)} \nonumber\\
{\bf 16} &=& {\bf (4,2,1) + (\bar{4},1,2)} \nonumber\\
{\bf \bar{16}} &=& {\bf(\bar{4},2,1) + (4,1,2)}
\end{eqnarray}
Now the vev of $A$ is along the (1,1,3) direction and hence the coupling
$PA\bar{P}$ makes all the fields in $(\bar{4},1,2)$ and (4,1,2) of $P$ 
and $\bar P$ massive. Since the vevs of $C$ and $\bar{C}$ are along the
right handed neutrino direction the only coupling which could potentially
give mass to the (4,2,1) of $P$ is of the form
\begin{equation}
<4,1,2>_{\bar{C}}(6,2,2)_{\bar{A}}(4,2,1)_{P}
\end{equation}
However because the 6 of $SU(4)$ is antisymmetric in its indices this 
fails to give mass to the lone $SU(2)_L$ doublet in $P$ which is 
uncharged under $SU(3)_C$. This doublet has the same quantum numbers
as $H_d$ of the standard model. 
In exactly the same manner a doublet in $\bar{P}$ which has the same
quantum numbers as $H_u$ is left massless. 
It is then a straightforward matter to
to mix the doublets in a pair of {\bf 10}s with those in the {\bf 16}s so 
that the light MSSM doublets emerge as linear combinations of the 
{\bf 10}s and the {\bf 16}s. 

In order to illuminate our our main ideas we construct an SO(10) model 
with the
following field content in the Higgs sector: one {\bf 54} (S); two
{\bf 45}'s ($A,\bar{A}$), three pairs of ${\bf 16\oplus \bar{16}}$ (denoted
by $C\oplus \bar{C}$, $P\oplus \bar{P}$ and $D\oplus \bar{D}$),
two {\bf 10}'s ($H_{1,2}$) and three singlets ($T, T',Y$). By analysing the
supersymmetry preserving conditions at the GUT scale we will show that
there is a vacuum where only $S, A, C\oplus \bar{C}, T, T'$ have vevs with 
the
new missing vev pattern for the $A$. We then show using the rest of the 
multiplets that one can implement our new doublet triplet splitting 
suggestion; the resulting MSSM doublets then lead to a 
pattern of fermion masses that is free of difficulties such as vanishing
CKM angle or bad quark lepton mass relations. 

\vskip0.3in

\section{The Symmetry Breaking Sector}

Let us split the superpotential for our example into three pieces:
$W= W_1 + W_2 + W_3$; $W_1$ being responsible for breaking SO(10) 
down to the MSSM and giving the desired SUSY
invariant pattern of vevs, $W_2$ causing the doublet triplet splitting 
and $W_3$
giving rise to the fermion masses and mixings. Let us first discuss
$W_1$:
\begin{eqnarray}
W_1~=~ \lambda S^3 + M_1S^2 + SA^2 + Y(C\bar{C}-M^2_2)+ M_3A^2\nonumber\\
+\bar{C}(A+T)D+C(A+T')\bar{D} + M_0D\bar{D}
\end{eqnarray}
We have scaled some dimensionless couplings in the above eqn. to 1 for 
simplicity.
Note that the remaining spinors and the adjoint have no role at this stage
since they can be self consistently assumed to have zero vev. Let us now 
assume
the ground state form for the various Higgs fields to be as follows:
$<A>=i\tau_2\otimes Diag (a_1, a_2, a_3, a_4, a_5)$, $<S>= {\bf 1}\otimes
Diag (s_1, s_2, s_3, s_4, s_5)$, $<C_{\nu^c}>=<\bar{C}_{\nu^c}>=v_R$
and $<T>=<T'>\neq 0$. The various F-term vanishings then imply the following
equations:
\begin{eqnarray}
3\lambda s^2_i + 2M_1 s_i + a^2_i +\gamma =0
\end{eqnarray}
where $\gamma$ denotes the Lagrange multiplier needed to guarantee the
tracelessness condition of the {\bf 54}-plet field. Next we have from 
$F_A=0$,
\begin{eqnarray}
a_i(M_3 + s_i)=0
\end{eqnarray}
The vanishing conditions for the other F-terms imply $\Sigma_i a_i= -<T>
=-<T'>$, $<D>=<\bar{D}>=0$, $<Y>=0$ and $v_R= M_2$. Let us therefore 
analyze the 
Eq. (4) and (5). Note first that Eq. (5) implies that either $a_i=0$ or
$s_i=-M_3$. This implies that there is always one solution for which
$a_{4,5}\neq 0$ and $s_4=s_5=-M_3$ with $a_{1,2,3}=0$ and $s_{1,2,3}$ 
arbitrary. Eq. (2) then implies that $s_{\alpha}$  ($\alpha=1,2,3$) satisfy
the equation
\begin{eqnarray}
3\lambda s^2_{\alpha}+2M_1 s_{\alpha} +\gamma=0
\end{eqnarray}
We can now choose $s_{1,2,3}$ all equal as a solution of the above 
equation. The tracelessness condition on $S$ then implies that 
$s_1 = 2M_3/3$. Then equation (6) determines $\gamma$ in terms of $M_1$
and $M_3$. Once $\gamma$ is known we can use equation (4) to determine 
$a_4=a_5\neq 0$. Hence all the vevs have been determined and we have 
established that the desired new missing vev pattern is a consistent 
solution of all vanishing F-term equations.

\section{Doublet-Triplet Splitting and the MSSM Doublets:}

In order to establish that our new vev pattern indeed leads to a useful
pattern of doublet-triplet splitting, let us write down $W_2$:
\begin{eqnarray}
W_2= f_1PA\bar{P} +f_2 C\bar{A}\bar{P} +f_3\bar{C}\bar{A}P\nonumber\\
+h_1CPH_1 +h_2\bar{C}\bar{P}H_2 +M_4H^2_1 +M_5 H^2_2 +M_6 \bar{A}^2
\end{eqnarray}
Since the all the fields in above equation other than $C,\bar{C}$ and $A$
have zero vev, we can easily write down the doublet mass matrix which is 
represented schematically as:
 \begin{eqnarray}
M_D=\left(\begin{array}{ccc} \bar{P}_u & H_{1u} & H_{2u}\end{array}\right)
\left(\begin{array}{ccc} 0 & <\bar{C}> & 0\\ <C> & 0 & M_4\\ 0 & M_5 & 
0\end{array}\right)\left(\begin{array}{c} P_d\\ H_{2d} 
\\H_{1d}\end{array}\right)
\end{eqnarray}
where the rows and columns denote the $SU(2)_L$ doublets in the {\bf 
10}'s and the spinors. From this equation we see that the above matrix 
has two zero eigenstates: $H_u\equiv c_1 \bar{P}_u + s_1 H_{2u}$ and
$H_d\equiv c_2 P_d + s_2 H_{1d}$. The remaining four doublets pick up
GUT scale mass.

Turning to the color triplet mass matrix, let us think in the SU(5)
submultiplet language. Note that since $C$ and $\bar{C}$ acquire vevs
along the SU(5) singlet direction, the $f_{2,3}$ couplings gives masses
to the SU(5) {\bf 10} and ${\bf \bar{10}}$ multiplets in the spinors
by combining them with the appropriate SU(5) $\bar{10}$ and {\bf 10} fields
from the $A$. Thus we only have to consider the mass matrix for the 
triplets from the SU(5) submultiplet {\bf 5} and ${\bf \bar{5}}$'s 
from the spinors and the $H_{1,2}$. The 
resulting mass matrix for them can be written schematically as:  
                  
\begin{eqnarray}
M_T=\left(\begin{array}{ccc} 
\bar{\xi}_1 & \bar{\xi}_{2} & 
\bar{\xi}_{\bar{P}}\end{array}\right)
\left(\begin{array}{ccc} 
M_4 & 0 & <C>\\ 0 & M_5 & 0 \\ 0 & <\bar{C}> & <A>
\end{array}\right)\left(\begin{array}{c} 
\xi_1\\ \xi_{2}
\\\xi_{P}\end{array}\right)
\end{eqnarray}
Clearly this matrix has no zero eigenvalues implying that all the color 
triplet states are now massive and at the GUT scale. Thus doublet 
triplet splitting has been successfully implemented. Since $H_{u}$ and $H_d$
emerge from different multiplets of SO(10), realistic up and down mass 
matrices can be constructed in a strightforward manner.

While this the most general scenario for doublet-triplet splitting,
variations on this theme can be constructed in particular applications.
Below we present one such case which helps in a natural understanding 
of why the top quark is so much heavier than the bottom quark. In the
subsequent section we show how the same model can provide us with a type II
seesaw realization which can generate a nonhierarchical pattern for 
neutrino masses.

\section{Understanding the top-bottom mass hierarchy}

In the SO(10) models quark lepton masses arise from two kinds of terms:
(i) renormalizable terms involving the {\bf 10} Higgses of the form
$\psi\psi H$ where the $\psi_a$ denotes the matter field spinor for 
generation $a=1,2,3$ and (ii) nonrenormalizable terms of several 
different kinds of which the ones of our interest are 
$\psi\psi\bar{ P}\bar{C}$
or $\psi\psi {P}C$ provided the MSSM doublets contain the doublet
pieces from the corresponding $H, P, \bar{P}$ after 
doublet-triplet splitting. Since the nonrenormalizable terms are 
suppressed by some higher mass scale such as the string scale or the 
Planck scale, their contribution to the quark lepton masses are 
naturally suppressed by a factor $M_U/M_{st}$ or $M_U/M_{Pl}$ which are 
respectively of order $1/20$ or $1/100$ compared to the mass terms that 
arise from the renormalizable terms.
It is then clear that if the $H_d$ of MSSM consists solely of the 
doublet in the spinor Higgs $P$ whereas the $H_u$ consists of the 
doublet in {\bf 10} with or without a contribution from $\bar{P}$, then 
the bottom quark mass is automatically suppressed compared to the top 
quark.                        
                                                                             
The particular appeal of this idea for the present doublet 
triplet splitting scenario is that our primary light doublet comes from 
the spinor Higgs fields and therefore all we have to do is to mix in 
{\bf 10} component in the light $H_u$ that to start with consists only of 
the doublet from the ${\bf \bar{16}}$. This is easily achieved if in 
the model discussed in the previous section we drop the $H_1$ field 
which means that the doublet mass matrix at the GUT scale looks like 
\begin{eqnarray}                                                                
M_D=\left(\begin{array}{cc} \bar{P}_u & 
H_{2u}\end{array}\right)   
\left(\begin{array}{cc} 0 & <\bar{C}> \\  0 & M_5 
\end{array}\right)\left(\begin{array}{c} P_d\\ 
H_{2d}\end{array}\right)                                                   
\end{eqnarray}                                                             
The light MSSM doublets are then given by $H_d\equiv P_{d}$ and 
$H_u\equiv \alpha H_{2u} + \beta \bar{P}_u$. This is the desired combination 
that naturally explains the top-bottom mass hierarchy as explained in the 
previous paragraph. Clearly, by appropriate choice of the various 
renormalizable and nonrenormalizable terms, we can obtain a realistic 
pattern of masses and mixings.

Two comments are in order here: (i) the model outlined above is more 
economical in the sense that we use only one {\bf 10} rather than two 
commonly used when the Dimopoulos-Wilczek  pattern is used; (ii) furthermore
in models that use DW pattern, it highly nontrivial to have the $H_d$
purely as coming from the {\bf 16}-Higgs field, whereas in our case it
is relatively straightforward.
 
\section{Type II seesaw formula and nonhierarchical neutrino masses}

Let us begin by reminding the reader that in SO(10) models where
the seesaw mechanism is implemented via the {\bf 126} 
Higgs fields, the neutrino mass has two contributions:
one which depends quadratically on the quark masses and a second one 
that depends quadratically on the electroweak scale. When the first one 
alone appears, we will call it type I seesaw formula. When both appear, 
we will call it type II seesaw formula\cite{goran,lee}. Since the 
second one is not related to the quark or lepton masses, it is in 
general nonhierarchical. Furthermore, the new contribution depends 
on the magnitude of the vev of the $SU(2)_L$ triplet field (to be denoted 
$\Delta_L$) in the {\bf 126} representation, which in turn depends on 
the mass of the $\Delta_L$. If we interested in a contribution to $m_{\nu}$
of order of an eV, one would expect a mass of the $\Delta_L$ to be of order 
$10^{12}$ GeV or so. This requires an intermediate scale unification.  

On the other hand, when there are no {\bf 126} representations in the 
theory as is apparently the case in string derived models\cite{dienes} 
small neutrino masses arise via the 
usual seesaw mechanism (type I) once we include nonrenormalizable terms of 
the form $\psi \psi \bar{C}\bar{C}/ M$ which will generate the mass 
matrix of the right handed Majorana neutrinos with arbitrary texture. To 
the best of our knowledge, this was considered to be the only 
contribution to 
the neutrino masses leading to belief that the neutrino masses in such 
models will be necessarily hierarchical.

We however find that if the MSSM doublet $H_u$ contains a piece from the 
${\bf \bar{16}}$ Higgs field, a new contribution
 to the light neutrino mass matrix can arise from the nonrenormalizable 
terms of the form $\psi\bar{P} \psi\bar{P}/M$. This then leads to the 
type II seesaw formula and hence in our case depending on the model, we 
can have either a hierarchical or a nonhierarchical mass pattern for 
neutrinos.

The magnitude of the nonhierarchical contribution depends on what is chosen 
for $M$. If we choose $M=M_{U}$ as would be the case in the minimal 
model, the nonhierarchical contribution can 
 be $\sim 10^{-3}$ eV. They can be larger if $M$ is a smaller mass 
than the GUT scale. The latter situation can happen for example in a model 
which has an extra singlets $Z_1$ and a superpotential of the form:
\begin{eqnarray}
W_4~=~\psi\bar{P}Z_1 +\lambda' Z^2_1 C\bar{C}/M_{Pl}
\end{eqnarray}
After integrating out the $Z_{1}$ field we get a value for the
scale $M= \lambda' v^2_R/M_{Pl}$ and is thus smaller than $M_U$ by at least
a couple of orders of magnitudes or so. They can therefore easily lead to
new contributions to the neutrino masses of order of an eV, as needed,
say for the hot dark matter component of the universe. A further 
implication is that in this model the unification is still a single stage
type (i.e. no intermediate scales).

A very simple modification of Eq. 11 can also lead to a maximal mixing 
between the $\nu_{\mu}$ and $\nu_{\tau}$. For instance if we choose
$W_4 = MT\bar{T} + T\psi_2\bar{P} +\bar{T}\psi_3{\bar{P}}$, then this 
leads to maximal mixing between the $\nu_{\mu}$ and $\nu_{\tau}$. The 
splittings can then arise out of the usual seesaw contributions. Of 
course to solve the solar neutrino problem, one must invoke the sterile 
neutrino.

To summarize this section,
even though the SO(10) model discussed in this paper does not have {\bf 126} 
Higgs fields, one can have significant new contributions to neutrino masses 
that alter the hierarchical pattern expected from the usual type I 
seesaw formula. The origin of these new contribution is linked to the new 
way of implementing the doublet-triplet splitting advocated in this paper.

In conclusion, we have presented a new mechanism for                       
doublet-triplet splitting in the supersymmetric 
SO(10) grand unification using a missing vev pattern that is different 
from the one currently most popular. We have provided an explicit realization
of this scheme using only renormalizable terms so that no 
unwanted states with intermediate scale masses arise and unification of 
couplings is maintained. We also show how the model can provide a natural
explanation of the top bottom mass hierarchy as well as a nonhierarchical
contribution to neutrino masses.

 This work is supported by the National Science Foundation 
under grant no. PHY-9802551.

{\it Note added in proof:} After this paper was submitted for 
publication, a paper by G. Dvali and S. Pokorski\cite{dvali} was brought to 
our attention. This paper has some similarity to certain aspects of our 
work in that it uses {\bf 45} vev patterns similar to ours and uses the
doublets in {\bf 16} as low energy doublets. However, the
ways to understand top bottom splitting is completely different.
The remarks in our paper about neutrino mass are also new.

\end{document}